\documentclass[fleqn,usenatbib]{mnras_tex_edited}

\usepackage{amssymb,amsmath,verbatim,mathtools,needspace,enumitem,etoolbox,graphicx,physics,microtype,natbib,url,hyperref,mathrsfs,bm,ulem,lipsum}
\usepackage{orcidlink}
\usepackage{siunitx}
\newcommand{\ssim}{\mathchar"5218\relax\,}

\renewcommand{\emph}[1]{\textit{#1}}

\newcommand{\approptoinn}[2]{\mathrel{\vcenter{
  \offinterlineskip\halign{\hfil$##$\cr
    #1\propto\cr\noalign{\kern2pt}#1\sim\cr\noalign{\kern-2pt}}}}}

\def\araa{{Ann.~Rev.~Astron.~Astrophys.}}
\def\apj{{Astrophys.~J.}}\def\apjl{{Astrophys.~J.~Lett.}}
\def\apjs{{Astrophys.~J.~Supp.}}
\def\ao{{Appl.~Opt.}}

\def\aap{{Astron.~Astrophys.}}

\def\mnras{{Mon.~Not.~R.~Astron.~Soc.}}             
\def\na{{New Astron.}}

\def\prd{{Phys.~Rev.~D}}

\def\prl{{Phys.~Rev.~Lett.}}

\def\ssr{{Space~Sci.~Rev.}}

\def\nat{{Nature}}

\def\physrep{{Phys.~Rep.}}

\def\lrr{{Living Rev. Relativ.}}
\def\cqg{{Classical~Quant.~Grav.}}

\newcommand{\milan}{Dipartimento di Fisica ``G. Occhialini'', Universit\'a degli Studi di Milano-Bicocca, Piazza della Scienza 3, 20126 Milano, Italy}
\newcommand{\infn}{INFN, Sezione di Milano-Bicocca, Piazza della Scienza 3, 20126 Milano, Italy}
\newcommand{\bham}{School of Physics and Astronomy \&	 Institute for Gravitational Wave Astronomy, University of Birmingham,\vspace{-0.05cm}\\$\;$Birmingham, B15 2TT, UK}

\title[Spin orientations of supermassive black-hole binaries]{
The Bardeen-Petterson effect, disk breaking, and the spin orientations of supermassive black-hole binaries 
}

\author[Steinle and Gerosa]{Nathan Steinle\thanks{\href{mailto:nsteinle@star.sr.bham.ac.uk}{nsteinle@star.sr.bham.ac.uk}}$\,$\orcidlink{0000-0003-0658-402X}$^{1}$,
Davide Gerosa$\,$\orcidlink{0000-0002-0933-3579}$^{2,3,1}$,
\medskip
\\
$^{1}$\bham\\
$^{2}$\milan\\
$^{3}$\infn\\
}

\date{}

\pubyear{\the\year{}}

\begin{document}
\label{firstpage}
\pagerange{\pageref{firstpage}--\pageref{lastpage}}
\maketitle

\begin{abstract}
Supermassive black-hole binaries are driven to merger by dynamical friction, loss-cone scattering of individual stars, disk migration, and gravitational-wave emission. Two main formation scenarios are expected. Binaries that form in gas-poor galactic environments do not experience disk migration and likely enter the gravitational-wave dominated phase with roughly isotropic spin orientations. Comparatively, binaries that evolve in gas-rich galactic environments might experience prominent phases of disk accretion, where the Bardeen-Petterson effect acts to align the spins of the black holes with the orbital angular momentum of the disk. However, if the accretion disk breaks alignment is expected to be strongly suppressed ---a phenomenon that was recently shown to occur in a large portion of the parameter space. In this paper, we develop a semi-analytic model of joint gas-driven migration and spin alignment of supermassive black-hole binaries taking into account the impact of disk breaking for the first time. Our model predicts the occurrence of distinct subpopulations of binaries depending on the efficiency of spin alignment. This implies
that future gravitational-wave observations of merging black holes could potentially be used to (i) discriminate between gas-rich and gas-poor hosts and (ii) constrain the dynamics of warped accretion disks. %
\end{abstract}

\begin{keywords}
accretion, accretion discs -- black-hole mergers -- gravitational waves -- quasars: supermassive black holes
\end{keywords}

\section{Introduction}
\label{sec:Intro}

The Laser Interferometer Space Antenna (LISA, \citealt{2017arXiv170200786A}) will observe mHz gravitational-waves (GWs), where supermassive binary black-holes (BHs) are a prime target (e.g. \citealt{2022arXiv220306016A,2021FrASS...8....7S}). Individual supermassive BHs are known to occupy the centers of most galaxies \citep{1995ARA&A..33..581K}. They originate as ``seeds'' from either the remnants of the first stars that populate the Universe or the direct collapse of large gas clouds, growing subsequently through accretion and hierarchical mergers \citep{2021NatRP...3..732V}. Binaries are understood to exist from observational evidence of past mergers \citep{2005LRR.....8....8M}, but the details of their pairing processes is one of the most outstanding problems in modern astrophysics. Observations of GWs by LISA present a unique opportunity of disentangling the cosmic evolutionary history of these objects. 

The evolution of supermassive BH binaries can be divided into four main phases according to the dominant mechanism of angular momentum loss \citep{1980Natur.287..307B,2014SSRv..183..189C}. %
In the first phase, the distance between the BHs 
decreases
from $\gtrsim 0.01$\,Mpc down to $\ssim 1$\,pc %
due to dynamical friction against the galactic stellar %
background. %
The second phase is dominated by the dynamical scattering of individual stars in the loss cone of the binary, decreasing the separation down to $\ssim 0.01$\,pc. If there is a sufficient reservoir of gas in the galactic host, viscous dissipation to the resulting accretion disk can further harden the binary. The final phase, dominated by the emission of GWs, drives the binary to merger. In recent decades, the apparent inability of loss-cone scattering to harden the binary into the GW-dominated phase (the so-called ``final-parsec problem,'' \citealt{2003AIPC..686..201M}), has been successfully explained by employing more realistic models of the galactic host \citep{2004ApJ...606..774P,2017MNRAS.464.2301G}. 

The occurrence of a phase of disk-driven migration provides two broad classes of evolutionary channels for the formation of supermassive BH binaries, depending on whether they merge in gas-rich or gas-poor galactic environments. The spin angular momenta of the BHs are predicted to be 
clean observables to distinguish between these two pathways~\citep{2007ApJ...661L.147B, 2008ApJ...684..822B, 2011PhRvD..83d4036S, 2013MNRAS.429L..30L, 2013ApJ...774...43M, 2013ApJ...762...68D, 2015MNRAS.451.3941G, 2021MNRAS.501.2531S}.  %
Although population-level inference on formation channels with LISA still needs to be perfected (see \citealt{2011PhRvD..83d4036S,2011CQGra..28i4018G,2021PhRvD.104h3027T}), spins are expected to provide a direct link to some of the key underlying astrophysical processes.  
This is analogous to the case of stellar-mass BH binaries observed by LIGO and Virgo, where the isolated and dynamical formation channels provide different predictions for the BH spin orientations (e.g.'s, \citealt{2021hgwa.bookE...4M,2022PhR...955....1M}).

Investigations into the evolution of supermassive-BH spin directions during the disk-driven inspiral can involve a vast array of different assumptions, such as prolonged
accretion with constant direction of the angular momentum \citep{2008ApJ...684..822B} vs. smaller isotropic accretion episodes \citep{2006MNRAS.373L..90K}, or more general cases considering different degrees of unisotropy in the fueling flow \citep{2013ApJ...762...68D,2014ApJ...794..104S}. 
The broad picture for BH binaries (e.g. \citealt{2007ApJ...661L.147B}) is that accretion in gas-rich galaxies largely align the BH spins while binaries that evolve in gas-poor galaxies retain isotropically distributed spin directions. 
Important caveats to this statement include potential alignment mechanisms during the dynamical friction phase prior to loss-cone scattering \citep{2010MNRAS.402..682D}. The spin magnitudes of the BHs can increase or decrease due to gas accretion depending on whether the disk is in prograde or retrograde orbit, respectively \citep{1999MNRAS.305..654K,2009MNRAS.399.2249P}. 

The evolution of a supermassive BH binary through a gas-rich environment has been studied extensively with both hydrodynamical simulations and semi-analytic models \citep{2014SSRv..183..189C}. As disk migration is driven by the dissipation of angular momentum from the binary to the circumbinary disk, the binary carves a cavity in the surrounding material resulting in the formation of smaller secondary disks around each BH, also known as ``minidisks'' (e.g. \citealt{2018ApJ...853L..17B}). A misaligned, spinning BH that is accreting gas from its secondary disk can induce Lense-Thirring precession, causing the disk to warp and eventually align with the BH spin ---a process commonly referred to as the Bardeen–Petterson effect \citep{1975ApJ...195L..65B,1978Natur.275..516R,1985MNRAS.213..435K,1998ApJ...506L..97N}. In a binary system, perturbations from the BH companion introduce an additional torque onto the %
secondary disks. The combination of Lense-Thirring and companion torques may lead to configurations where the disk breaks into distinct sections or rings. This is the so-called ``critical obliquity" phenomenon first identified by \cite{2014MNRAS.441.1408T} and explored at length by \cite{2020MNRAS.496.3060G}.  
For more context on the Bardeen-Petterson effect and disk breaking see 
\cite{
2000MNRAS.315..570N,
2012ApJ...757L..24N,
2013MNRAS.434.1946N,
2012MNRAS.421.1201N,
2015MNRAS.449.1251D,
2015MNRAS.448.1526N,
2018MNRAS.476.1519D,
2016MNRAS.455L..62N,
2022MNRAS.509.5608N,
2020MNRAS.495.1148D,
2021ApJ...909...81R,
2021MNRAS.507..983L}.

In this work, we assess the impact of disk-assisted spin alignment on future observations of merging supermassive BH binaries. %
In particular, we target the distinguishability of sources formed in gas rich vs. gas poor environments using future spin measurements. We model the BH evolution in gas-rich environments assuming that the binary was previously hardened by dynamical friction and loss-cone scattering of individual stars. We then capture disk migration and spin alignment using the one-dimensional approach by \cite{2020MNRAS.496.3060G}, which include both the non-linear effects of the fluid viscosities \citep{1999MNRAS.304..557O,2013MNRAS.433.2403O} as well as the perturbations induced by the binary companion. This allows us to consider, for the first time in a supermassive-BH binary formation model, the effect of the disk critical obliquity and its impact on the broader population of GW sources.  %
Our disk modeling serves as an initial condition for the subsequent phase of the binary evolution where GW emission dominates, which we capture with a 
post-Newtonian scheme specifically designed to bridge large astrophysical separations to the last orbits before merger~\citep{2016PhRvD..93l4066G}. %
Leveraging our two-step model (Bardeen-Petterson effect and post-Newtonian evolution) we find that disk breaking has a critical impact on the supermassive-BH spin-alignment process in gas-rich galaxies. %

This paper is organized as follows. In Sec.~\ref{sec:Meth}, we present the adopted model of disk migration and accretion. In Sec.~\ref{sec:Results}, we discuss the evolution of the spin orientations of individual binaries, the emergence of subpopulations of binaries with distinct spin orientations, and the dependence of these subpopulations on the various parameters that set the underlying disk physics. In Sec.~\ref{sec:Discussion}, we conclude with a summary and discussion of implications for LISA observations.

\section{Model}
\label{sec:Meth}

While full cosmological simulations are necessary to investigate the supermassive-BH pairing processes in detail, we argue the essential ingredients setting the spin orientations can be encapsulated with relatively simple semi-analytical prescriptions. %

\subsection{Initialization of disk migration}
\label{initialization}

Dynamical friction and loss-cone scattering are thought to %
weakly affect the BH spin directions %
on long timescales \citep{2012PhRvD..86j2002M}, however this comes with significant uncertainties. We assume that the BHs are paired with initially isotropic spin directions and that alignment mechanisms are avoided during the dynamical-friction phase. This implies that binaries evolving in gas-poor environments will enter the GW-driven phase with isotropically distributed spins which is preserved to high accuracy through the post-Newtonian inspiral \citep{2007ApJ...661L.147B,2015PhRvD..92f4016G}. These binaries will therefore be seen by LISA %
 with isotropic spins directions.
 
Conversely, for binaries evolving in the gas-rich channel, the accretion disk introduces a preferential direction that breaks isotropicity. BH spins with initially isotropic directions are subject to the Bardeen-Petterson effect and may align with the orbital angular momentum of the disk. For simplicity, in the following we assume that the circumbinary disk (which is responsible for the BH migration) and the outer edges of secondary disks (which are responsible for the spin alignment) share the same orientation \citep{1999MNRAS.307...79I}, the latter being fed by the former (e.g. \citealt{2014ApJ...783..134F}). We also assume that the binary orbit lies in the plane of the circumbinary disk. %

To obtain the initial separation for disk migration, we define the hardening timescale on which angular momentum is lost via loss-cone scattering of single stars by a parameterized power-law, %
\begin{align}\label{E:HardeningTimescale}
\begin{aligned}
    t_{\rm h} = t_{\rm s} \left( \frac{r}{R_{\rm s}} \right)^\delta \,
\end{aligned}
\end{align}
where $t_{\rm s}$ and $R_{\rm s}$ are scaling parameters and $r$ is the binary separation. The broad expectation \citep{1996NewA....1...35Q} is that $t_{\rm h} \sim \sigma/\rho Gr$, where $G$ is the gravitational constant, and $\rho$ is the density profile and $\sigma$ is the 1-dimensional velocity dispersion of the stellar background. We take a fiducial model with  $\delta =-1$, $t_{\rm s} = 10$ Myr, and $R_{\rm s} = 0.1$ pc  (cf. \citealt{2017MNRAS.464.3131K}). %

Similarly, we parameterize the inspiral timescale during the disk-driven migration as \citep{2020MNRAS.496.3060G}
\begin{align}\label{E:InspiralTimescale}
\begin{aligned}
    t_{\rm in} = \frac{t_{\rm b}}{f_{\rm T}} \left( \frac{r}{R_{\rm b}} \right)^{\gamma}\,,
\end{aligned}
\end{align}
where $\gamma$ is a free parameter that is of order unity, $t_{\rm b}$ and $R_{\rm b}$ are scaling factors, and $f_{\rm T}$ is the Eddington fraction of the circumbinary disk. In our fiducial model, we assume that $R_{\rm b} = 0.05$ pc, $f_{\rm T} = 0.1$, and $t_{\rm b} = 1$ Myr \citep{2003MNRAS.339..937G,2005ApJ...630..152E,2009ApJ...700.1952H,2017MNRAS.469.4258T,2017MNRAS.464.3131K,2019MNRAS.482.4383F}.
Simple arguments based on type-2 planetary migration \citep{1995MNRAS.277..758S,2013ApJ...774..144R,2015MNRAS.451.3941G} for disk-dominated systems %
suggest $\gamma=3/2$, which we take as our fiducial value (see also \citealt{2009ApJ...700.1952H}).

The transition $r_{\rm i}$ between the star-dominated and the gas-dominated regimes is obtained by equating Eq.~(\ref{E:HardeningTimescale}) and Eq.~(\ref{E:InspiralTimescale}). One gets
\begin{align}\label{E:InitialSepScaling}
\begin{aligned}
    r_{\rm i} =  \left( f_{\rm T}  \frac{t_{\rm s}}{t_{\rm b}}  \frac{R_{\rm b}^{\gamma}}{R_{\rm s}^{\delta}} \right)^{1/(\gamma - \delta)} \,,
\end{aligned}
\end{align}
such that $r_{\rm i} = 0.066$ pc %
for our fiducial setting.

\subsection{Warped disk structure}

As the binary inspirals, each BH accretes from its own disk. %
In particular, the system is defined by a timescale separation \citep{2013ApJ...774...43M,2020MNRAS.496.3060G}: the inner region of the accretion disk 
aligns with the BH spin on a timescale shorter than the time it takes for the BH spin to align with the outer region of the disk,
which is itself a shorter timescale compared to the time it takes for the BH mass and spin magnitude to change appreciably. %
The spin alignment process can thus be modeled quasi-adiabatically as a series of steady-state solutions to the disc evolution equations, while assuming the BH masses and spin magnitudes are constant  [see Eq.~(43) of \cite{2020MNRAS.496.3060G}]. 

We approximate the disk mass and angular-momentum profile using the iterative scheme put forward by \cite{2020MNRAS.496.3060G} (see their Sec. 3.3 for a full description of our framework). As an initial guess for our iterative scheme, we first solve the steady-state conservation equations for the disk structure \citep{1992MNRAS.258..811P} assuming linear viscosity coefficients 
[see Eq. (32) in \cite{2014MNRAS.441.1408T}].
Both Lense-Thirring precession and the perturbation of the BH companion are %
treated as external torques. We then estimate the non-linear viscosity profiles for that fixed disk structure using the locally isothermal theory by \cite{2013MNRAS.433.2403O} [see their Eqs.~(95-97)]. 
The resulting viscosities are then plugged back into the conservation equations [see Eqs. (34, 35) in \cite{2020MNRAS.496.3060G}], 
and the procedure is iterated until convergence. This allows us to compute the steady state solution of the disk  very efficiently using a simpler boundary-value solver instead of tackling the complete set of partial differential equations until relaxation as done by, e.g., \cite{2014MNRAS.441.1408T}.

The disk solution depends on four parameters:
\begin{enumerate}[leftmargin=*]
\item First, one needs to specify the kinematic viscosity coefficient $\alpha$ \citep{1973A&A....24..337S}. This is one of the key parameters that most affects our findings because it sets the portion of the parameter space where disks can break. Our fiducial runs are presented with $\alpha=0.2$.
\item The spectral index of the viscosity profile $\beta$ (\citealt{2007MNRAS.381.1617M}) instead has a negligible impact on the overall phenomenology \citep{2020MNRAS.496.3060G}. For our fiducial model we consider globally isothermal disks, i.e. $\beta=3/2$.
\item The tidal parameter \citep{2020MNRAS.496.3060G} %
\begin{align}\label{E:kappa}
    \kappa & \simeq 0.66 \left( \frac{m}{10^7\,{\rm M}_{\odot}} \right)^2 \left( \frac{\chi}{0.5} \right)^2 \left( \frac{m_{\rm c}}{10^7\,{\rm M}_{\odot}} \right) \left( \frac{r}{0.1\,\rm pc} \right)^{-3} 
    \notag \\
    & \times \left( \frac{H/R}{0.002} \right)^{-6} \left( \frac{\alpha}{0.2} \right)^{-3} \left[ \frac{\zeta}{1/(2\times0.2^2)} \right]^{-3}\,,
\end{align}
 sets the importance of the external torque at large radii; see also \cite{2014MNRAS.441.1408T} for a related parametrization. 
 In a nutshell, the disk equations can be reduced to a one-parameter family of solutions according to $\kappa = (R_{\rm tid}/R_{\rm LT})^{-7/2}$, where $R_{\rm tid}$ and $R_{\rm LT}$ are the disk radii where the companion tidal and Lens-Thirring torques, respectively, mostly affect the warp profile [cf. \cite{2009MNRAS.400..383M} and Eqs.~(13-22) in \cite{2020MNRAS.496.3060G}]. 
 In particular, larger (smaller) values of $\kappa$ corresponds to disks solutions that are more (less) perturbed by the binary companion, corresponding to a larger (smaller) warp amplitude.
 In the above equation, $m$ is the mass of the aligning BH, $\chi = c|\mathbf{S}|/Gm^2$ is the dimensionless spin magnitude 
 of the aligning BH with angular momentum $\mathbf{S}$, 
 $m_c$ is the mass of the BH companion, $r$ is the binary separation, $H/R$ is the aspect ratio at the reference radius where the viscosities are quoted \citep{2007MNRAS.381.1617M,2009MNRAS.400..383M}, and $\zeta = \zeta(\alpha)$ is the ratio of the vertical to horizontal viscosity in the small-warp limit  (with $\zeta\propto 1/2\alpha^2$ for $\alpha\to0$, \citealt{1983MNRAS.202.1181P,1999MNRAS.304..557O}).  %
For our fiducial model, we take $H/R = 0.002$ \citep{2009ApJ...700.1952H}. %

\item Finally, one needs to specify the misalignment of the angular momentum of the outer edge of the circum-BH disk $\theta\in[0,
\pi]$ with respect to the BH spin. %
\end{enumerate}

In some regions of this parameter space, 
the system can reach the critical obliquity \citep{2014MNRAS.441.1408T,2020MNRAS.496.3060G} where the underlying boundary-value problem does not admit solutions.
This corresponds to disk breaking as confirmed by recent 3D hydrodynamical simulations \citep{2022MNRAS.509.5608N}. %
In general, %
the disk breaks for low values of $\alpha$ (because some of the viscosity coefficients can become negative, \citealt{2018MNRAS.476.1519D}), large values of $\kappa$ (because the disturbance from the companion causes a drop in the surface density), and misalignments $\theta$ close to $90^\circ$ (because the warp profile becomes sharper). The condition for the disk to break is  $\theta_{\rm crit}<\theta<\pi - \theta_{\rm crit}$, where the threshold $\theta_{\rm crit}$ 
depends mostly on $\alpha$ and $\kappa$. 
As the binary migrates, the parameter $\kappa\propto r^{-3}$ increases, implying that disks can \emph{become} critical and break while the migration and alignment processes are taking place. %

Once the disk profile has been solved for, the time variation of the spin misalignment angle $d\theta/ d t$ can be computed by integrating the Lense-Thirring torque density \citep{2009MNRAS.399.2249P}. For cases where the disk breaks, the spin evolution is highly uncertain but recent hydrodynamical simulations by \cite{2022MNRAS.509.5608N} seem to indicate that alignment is suppressed (although it is important to note that the length of their runs are much shorter than the BH inspiral timescale). This reflects one's intuition that a broken disk interrupts, but does not prevent, the flow of angular momentum. In the absence of a more accurate parametrization, we assume that  the spin does not evolve at all after the disk breaks. This corresponds to setting $d \theta / dt=0$ at all times beyond criticality. See section~\ref{sec:Discussion} for a discussion about this assumption.

Disk criticality is a key new feature of our model and, as we explore at length below, it imprints a distinct signature in the expected distribution of spin misalignments.

\subsection{Quasi-adiabatic evolution}

Disk-driven migration is implemented with a quasi-adiabatic approximation. At each timestep, we estimate $d \theta / d t$ from the steady-state solution and evolve the separation according to $\dd r /\dd t =  -r / t_{\rm in}$, cf. Eq.~(\ref{E:InspiralTimescale}). We thus compute $\dd\theta/dt = \dd \theta/\dd t \times \dd t/\dd r$ which we solve numerically to obtain the evolution $\theta(r)$ of the spin angle as the binary migrates. 

While the efficiency of the migration process depends on the accretion rate of the circumbinary disk $f_{\rm T}$, spin alignment is set by the accretion properties of the smaller, 
secondary disks $f_{1,2}$. The key prescription entering our model here is that of ``differential accretion'' \citep{2014ApJ...783..134F,2015MNRAS.451.3941G,2020MNRAS.498..537S}, namely the expectation that 
material accretes preferentially onto the less massive BH because it orbits closer to the edge of the cavity in the circumbinary disk. We employ two simplifying assumptions \citep{2014ApJ...783..134F}:
\begin{enumerate}[leftmargin=*]
\item All material reaches the binary, i.e., $f_1+f_2=f_{\rm T}$. 
\item Differential accretion scales linearly with the BH masses, i.e.,  $f_2/f_1 = m_1/m_2$. 
\end{enumerate}
Improving upon these prescriptions with calibration on hydrodynamical simulations is an  interesting avenue for future work.  \cite{2016MNRAS.460.1243R} found a substantial pile-up of material at the edge of the cavity for very thin disks, thus  suggesting an additional $H/R$ dependence which is not captured by our model. \cite{2019MNRAS.485.1579K} presented a more elaborate differential-accretion prescription that could also be investigated. %

Within these assumptions, the global behavior of the $\theta(r)$ profiles is governed by a single dimensionless parameter \citep{2020MNRAS.496.3060G}
\begin{align}\label{E:Omega}
\begin{aligned}
    \omega &\simeq \left(0.54\times10^{0.55\gamma}\right) \left( \frac{m}{10^7\,{\rm M}_{\odot}} \right)^{-1 + 2\gamma/3} \left( \frac{\chi}{0.5} \right)^{2(\gamma - 1)/3} \\ 
    & \times \left( \frac{m_{\rm c}}{10^7\,{\rm M}_{\odot}} \right)^{1+\gamma/3} \left( \frac{R_{\rm b}}{0.05\,\rm pc} \right)^{-\gamma} \left( \frac{t_{\rm b}}{10^6\,\rm yr} \right) \\
    & \times \left( \frac{H/R}{0.002} \right)^{-2(\gamma + 1/3)} \left( \frac{\alpha}{0.2} \right)^{-\gamma - 1/3} \left[ \frac{\zeta}{1/(2\times0.2^2)} \right]^{2/3 - \gamma} \,\,.
\end{aligned}
\end{align}
In particular,
$\omega\propto t_{\rm in}/t_{\rm align}$ 
is related to the ratio of the timescales over which migration [$t_{\rm in}$, see Eq.~(\ref{E:InspiralTimescale})] and spin-alignment [$t_{\rm align}$, see Eq. (40) of \cite{2020MNRAS.496.3060G}] occurr
The parameter $\omega$ therefore acts much like the ``speed'' of the $\theta(r)$ evolution. Systems with $\omega \gg 1$ are expected to align quickly while systems with $\omega \ll 1$ do not have enough time to align during the disk-driven regime of the binary inspiral.

For a binary BH, we denote the masses of the two objects with $m_1\geq m_2$ and the dimensionless spin magnitudes with $\chi_{1,2} \in [0,1]$. The equations written thus far need to be used with $(m=m_1, m_c=m_2, \chi=\chi_1)$ when considering the more massive BH, and conversely $(m=m_2, m_c=m_1, \chi=\chi_2)$ when considering the less massive BH. This results in two ``companion'' parameters, $\kappa_1$ and $\kappa_2$, and two ``speed'' parameters, $\omega_1$ and $\omega_2$. %

\begin{figure}
\centering
\includegraphics[width=0.4825\textwidth]{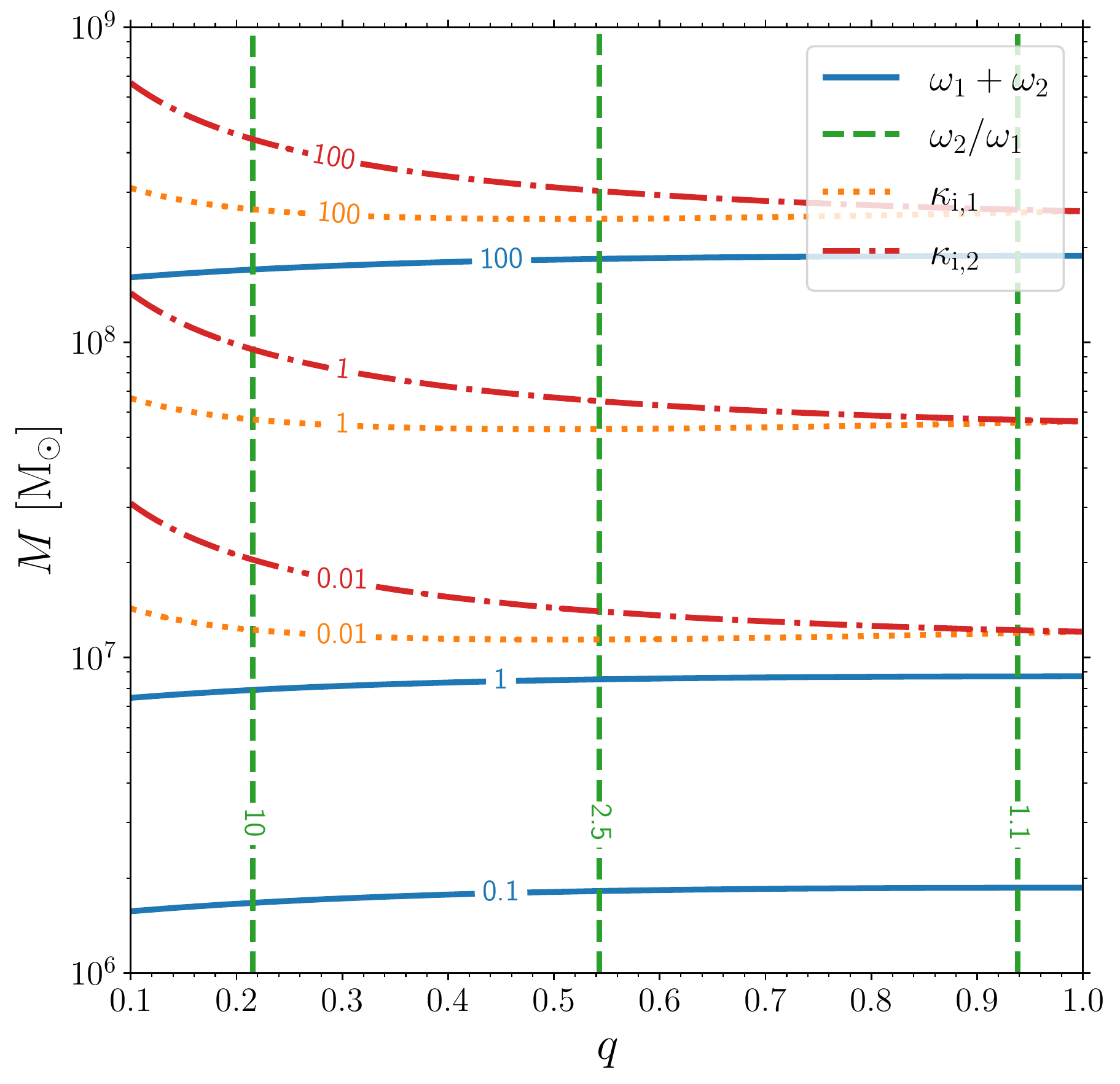}
\caption{Key parameters that govern the effectiveness of the spin-alignment process in supermassive BH binaries. We consider a set of BH binaries with different total mass $M$ and mass ratio $q$. Spin magnitudes are fixed to $\chi_1=\chi_2= 0.1$; 
disk parameters are set to their fiducial values. Solid blue and dashed green curves indicate the sum and ratio of the ``speed'' parameters $\omega_{1,2}$, respectively. The orange dotted and red dot-dashed curves indicate the values of the ``companion'' parameters $\kappa_{\rm i,1}$ and $\kappa_{\rm i,2}$ evaluated at the separation at which we initialize disk migration (cf. Eq.~\ref{E:InitialSepScaling}). Labels $1$ and $2$ refer to the more and less massive BH, respectively.%
}

\label{F:Contours}%
\end{figure}

The sum $\omega_1 + \omega_2$  and ratio 
\begin{align}\label{E:OmegaRatio}
\frac{\omega_2}{\omega_1} = 
\left(\frac{m_2}{m_1}\right)^{-2 +\gamma/3}  \left(\frac{\chi_2}{\chi_1}\right)^{2(\gamma-1)/3} \,,
\end{align}
turn out to be useful parametrizations for determining the behavior of the spin orientations in a binary system.
Figure~\ref{F:Contours} shows contours of these quantities as functions of the BH-binary total mass $M=m_1+ m_2$ and mass ratio $q=m_2/m_1$ assuming $\chi_1=\chi_2= 0.1$ 
and our fiducial disk parameters. The sum $\omega_1 + \omega_2$ increases with increasing $M$ and is $\gtrsim 1$ for $M \gtrsim 10^7$ M$_{\odot}$ where the efficiency of alignment of the BH spins is greatest. The ratio $\omega_2/\omega_1$ is independent of $M$ and proportional to $q$. For our fiducial values $\chi_1=\chi_2$ and our fiducial value $\gamma=3/2$, Eq.~(\ref{E:OmegaRatio}) yields $\omega_2/\omega_1 = q^{-3/2}$
and approaches unity in the limit $q\to1$. For these equal spin systems one has $\omega_2/\omega_1 > 1$, implying that alignment of the secondary is always faster than that of the primary. 
Instead, if $\chi_1 > \chi_2$ then $\omega_2/\omega_1 < 1$ is also possible. 

Figure~\ref{F:Contours} also shows the values of the companion parameters $\kappa_{1,2}$ evaluated at the separation  $r_{\rm i}$ where disk-assisted inspiral begins. For $\chi_1=\chi_2$ as assumed here, one has $\kappa_{\rm i,1} = \kappa_{\rm i,2}$  for $q\to1$, implying that both BHs are equally likely to begin their disk migration at a critical configuration. For BHs with $M \lesssim 10^8$ M$_{\odot}$, which are the likely targets for LISA observations, one has $\kappa_{\rm i,1} \lesssim 1$ and $\kappa_{\rm i,2} \lesssim 1$, implying a somewhat lower fraction of systems with broken disks (c.f Fig.~8 of \citealt{2020MNRAS.496.3060G}).
The contours in Fig.~\ref{F:Contours} are computed with a fiducial viscosity $\alpha = 0.2$. The contours of $\omega_1 + \omega_2$ and $\omega_2/\omega_1$ are largely insensitive to variations in $\alpha$. On the other hand, setting the kinematic viscosity to $\alpha = 0.1$ ($0.3$) provide values of $\kappa_{\rm i,1}$ and $\kappa_{\rm i,2}$ that are larger (smaller) by about a factor $\lesssim 2$ compared to the fiducial case.

\subsection{Disk decoupling and relativistic spin evolution}

In the late inspiral, the binary evolution becomes driven by GW emission. 
If gas is abundant, disk migration proceeds until the viscous timescale is smaller than gravitational radiation-reaction timescale. The transition separation where the binary and the disk decouple is given by \citep{2012PhRvL.109v1102F,2014PhRvD..89f4060G},
\begin{align}\label{E:DecoupSep}
\begin{aligned}
    r_{\rm decoup} &= \num{3e-4} \left( \frac{m_1+m_2}{\num{2e7}\,\rm M_\odot} \right) \left[ \frac{4m_1 m_2}{(m_1+m_2)^2} \right]^{2/5} \\
    & \quad \times \left( \frac{H/R}{0.002} \right)^{-4/5} \left( \frac{\alpha}{0.2} \right)^{-2/5} {\rm pc}\,.
\end{aligned}
\end{align}
In the opposite scenario where gas is insufficient, the largest possible separation resulting in a successful merger is given by 
\begin{align}
r_{\rm Hubble} &= 0.014 \left( \frac{m_1+m_2}{\num{2e7}\,\rm M_\odot} \right)^{3/4}\left[ \frac{4m_1 m_2}{(m_1+m_2)^2} \right]^{1/4} {\rm pc}\,,
\end{align}
at which the time to merger \citep{1963PhRv..131..435P} 
equals the age of the Universe. 

We find that $r_{\rm Hubble}/r_{\rm decoup}\gtrsim 10$ 
for total masses $M\gtrsim 10^{6}$~M$_{\odot}$. 
For $M\gtrsim 10^{7}$~M$_{\odot}$, the spin alignment process is %
very efficient (i.e., $\omega_1+\omega_2 \gtrsim 1$, cf. Fig.~\ref{F:Contours}) and the spins are aligned before the binary reaches $r_{\rm Hubble}$. This implies that
considering $r_{\rm Hubble}$ rather than $r_{\rm decoup}$ as the end point of disk evolution in our model is %
only relevant for a narrow portion of the parameter space. 
For simplicity, all our binaries are initialized at $r_{\rm i}$ from Eq.~(\ref{E:InitialSepScaling}) and halted at $r_{\rm decoup}$ from Eq.~(\ref{E:DecoupSep}).

Following the gas driven migration phase, %
we evolve the binary through its relativistic inspiral %
down to the separation 
\begin{equation}
r_{\rm GW} = 10 \frac{G(m_1+m_2)}{c^2}=9.6 \times 10^{-6}    \left( \frac{m_1+m_2}{\num{2e7}\,\rm M_\odot} \right) {\rm pc}
\label{rgw}
\end{equation} %
where they enter the sensitivity band of LISA and become detectable in GWs. We use post-Newtonian equations of motion averaged over the orbital period, %
as implemented by \cite{2016PhRvD..93l4066G}. We assume that the direction of the angular momentum of the binary is the same as that of the circumbinary disk, such that the spin-disk angles $\theta_i$ inherited from  the gas-driven phase are equal to the spin-orbit misalignments at the start of the relativistic inspiral. 

The aligned effective spin $\chi_{\rm eff} = (m_1 \chi_1 \cos\theta_1 + m_2\chi_2 \cos\theta_2)/(m_1+m_2)$ %
of a BH binary is a constant of motion at second post-Newtonian order \citep{2008PhRvD..78d4021R}. This implies that binaries can only evolve along straight, inclined lines in the $(\cos\theta_1-\cos\theta_2)$ plane \citep{2010PhRvD..81h4054K}. The extent of the spanned segment can be computed semi-analytically using the spin-precession solutions\footnote{In their notation, this corresponds to evaluating $\theta_1$ and $\theta_2$ at $S_{\pm}$ given the final value of $J$ resulting from the orbit-averaged integration.} by \cite{2015PhRvL.114h1103K} and \cite{2015PhRvD..92f4016G} evaluated at $r_{\rm GW}$.

\section{Results}
\label{sec:Results}

The input parameters of our model for the evolution of binary BH spin orientations are
\begin{enumerate}[leftmargin=*]
\item $\alpha$, $\beta$, $H/R$, and $f_{\rm T}$ for the disk properties;
\item $t_{\rm b}$, $R_{\rm b}$, $t_{\rm s}$, $R_{\rm s}$, and $\gamma$ for the timescale prescriptions;
\item $m_1$, $m_2$, $\chi_1$, and $\chi_2$ for the two BHs.
\end{enumerate}
The most crucial derived quantities are $\kappa_{1,2}$, which parameterize the effect of the companion on the accretion of each BH, and $\omega_{1,2}$, which parameterize the relative effects of alignment and inspiral. Unless specified otherwise, results are reported assuming the following fiducial values: 
$\alpha=0.2$, 
$\beta=3/2$
$H/R=0.002$, 
$f_{\rm T}=0.1$,
$t_{\rm b} = 10^6\,{\rm yr}$,
$R_{\rm b}=  0.05\,{\rm pc}$,
$t_{\rm s} = 10^7\,{\rm yr}$,
$R_{\rm s} = 0.1\,{\rm pc}$,
and 
$\gamma = 3/2$,

\subsection{Inspiral evolution}
\label{subsec:ResSpecific}

Our model of binary BH spin evolution returns three possible outcomes for each BH:

\begin{enumerate}[leftmargin=*]
    \item For the cases where the disk does not break, the Bardeen-Petterson effect is very efficient and the spin aligns almost completely.
        
    \item 
If the system is initialized in a configuration that is already past criticality (i.e. the disk is broken at $r_{\rm i}$), the spin does not evolve and maintains its initially isotropic orientation.

    \item %
    Disks that 
    have a stable configuration at the beginning but reach criticality during the gas-driven migration result in 
    partially aligned BH spins.
\end{enumerate}
While the first scenario corresponds to the broad conclusion reached by e.g. \cite{2007ApJ...661L.147B} and \cite{2013ApJ...774...43M}, the inclusion of disk breaking in our model provides 
for different, and potentially distinguishable, subpopulations. 
These three cases yield distinct, potentially observable GW signatures. %

\begin{figure*}
\centering
\includegraphics[width=0.88 \textwidth]{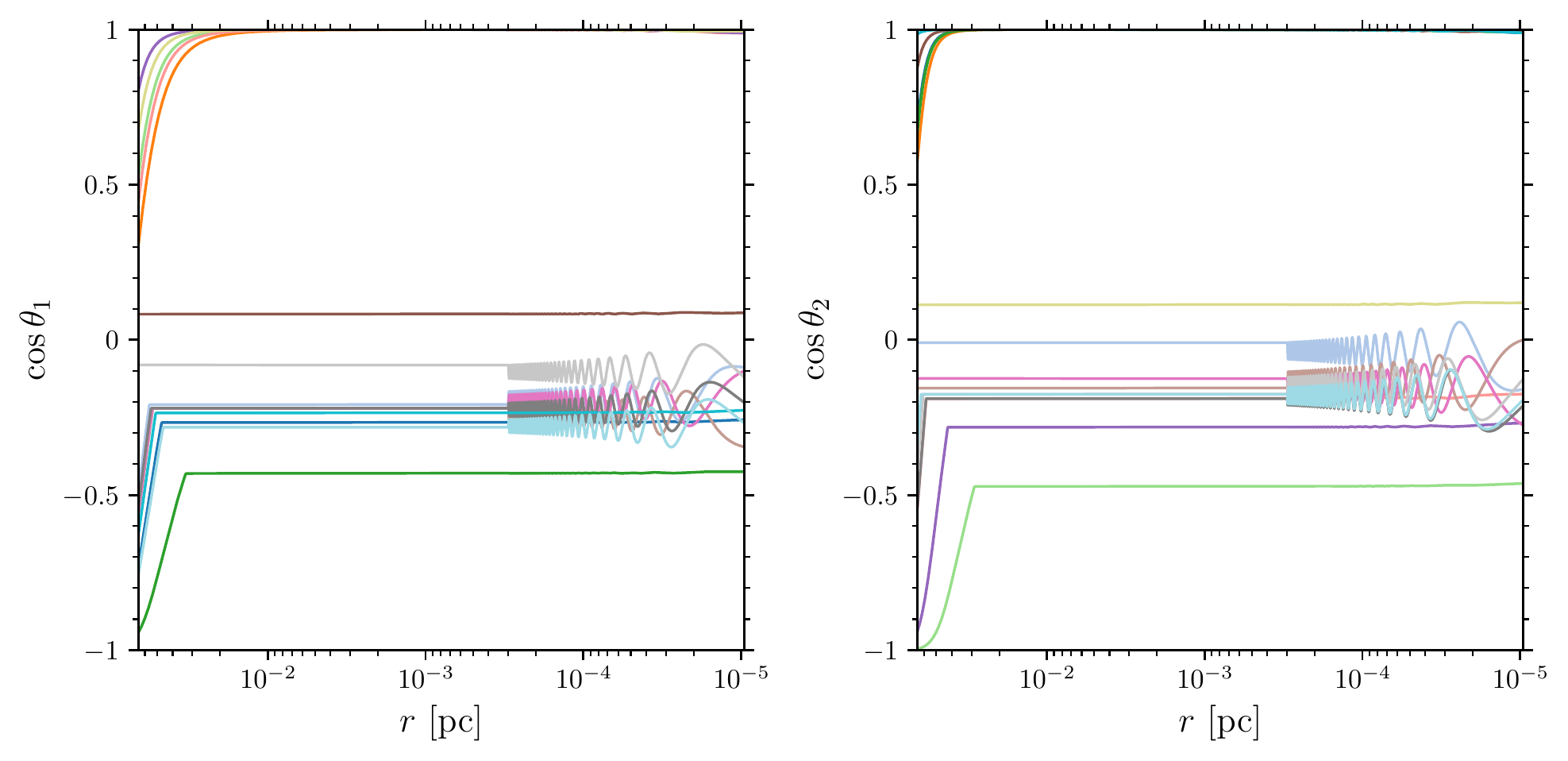}
\caption{The inspiral evolution of the spin orientations of the primary (left panel) and secondary (right panel) BHs in binaries with our fiducial disk parameters, total mass $M = 2\times10^7$ M$_{\odot}$, mass ratio $q = 0.8$, and spin magnitudes $\chi_1 = \chi_2 = 0.1$. Each pair of BHs that form a binary have a unique color. The initial spin orientations are isotropically distributed (i.e. flat in $\cos\theta_{1,2})$. While the Bardeen-Petterson effect tends to align the BH spins ($\cos\theta\to 1$), alignment ceases when the disks break and the spin directions remain constant. The sharp feature at $r=r_{\rm decoup} \simeq 0.0003~{\rm pc} \simeq 300~G M /c^2$ corresponds to the transition between the disk-migration and GW-dominated phases. 
} \label{F:Inspiral}%
\end{figure*}

\begin{figure}
\centering
\includegraphics[width=0.48\textwidth]{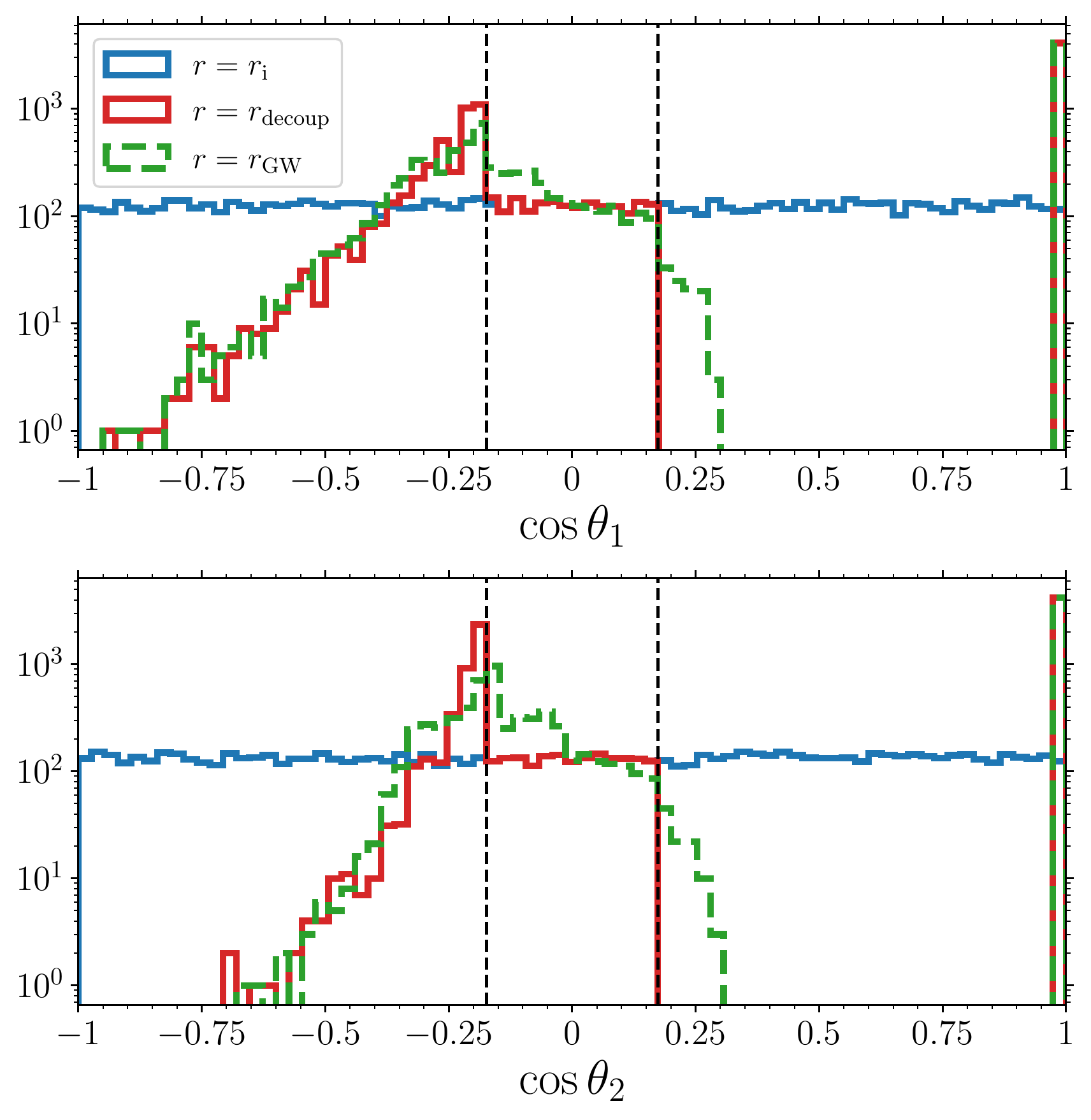}
\caption{Distributions of the spin-orbit misalignments of the primary (top) and secondary (bottom) BHs at the initial separation ($r=r_{\rm i}$, blue), at the end of disk evolution ($r=r_{\rm decoup}$, red), and close to merger ($r=r_{\rm GW}$, green) for supermassive binary BHs that evolve with our fiducial initial parameters, 
total mass $M = 2\times10^7$ M$_{\odot}$, mass ratio $q = 0.8$, and spin magnitudes $\chi_1 = \chi_2 = 0.1$. The initial spin orientations are isotropically distributed, i.e., uniform in cosine. The black, dashed vertical lines depict the values of the minimum and maximum critical angles, which are approximately equal for the primary and secondary BHs in these distributions because $\kappa_{\rm i,1} \simeq \kappa_{\rm i,2}$. 
} \label{F:Hist}%

\end{figure}

Figure~\ref{F:Inspiral} shows the evolution of the BH-spin orientations assuming our fiducial disk parameters and binaries with total mass $M = 2\times10^7$ M$_{\odot}$, mass ratio $q = 0.8$, 
spin magnitudes $\chi_1=\chi_2=0.1$, 
and isotropic spin directions. 
As these binaries only differ by the initial spin directions, all of the primary and secondary BHs evolve with the same value of  $\omega_1 \simeq 1.5$ and  $\omega_2 \simeq 2$ (cf. Fig.~\ref{F:Contours}). This implies that both BHs in each binary experience alignment moderately quickly with slightly faster alignment for the secondary than the primary since $\omega_2/\omega_1 > 1$. For these initial parameters, the region of the parameter space occupied by broken disks is approximately the same for both BHs as $\kappa_{1,\rm i} \simeq 0.05 \sim \kappa_{2,\rm i} \simeq 0.04$. When a BH reaches a critical angle, either because it is initialized %
at or it encounters one %
during the inspiral, we assume the spin direction remains constant through disk migration. The gas-driven phase ends at $r=r_{\rm decoup} \simeq 0.0003~{\rm pc} \simeq 300~G M /c^2$, %
after which the binary evolves under gravitational radiation reaction. The sharp transition $r_{\rm decoup}$ is an artifact of neglecting all relativistic effects before decoupling and we expect it to be smoother in more realistic %
models. While the mass ratio $q = 0.8$ allows for modest variation of the spin orientations through the GW dominated phase, systems with at least one spin aligned from disk migration do not experience significant spin evolution \citep{2015PhRvD..92f4016G}.

The histograms in Fig.~\ref{F:Hist} show three snapshots of the spin evolution of the primary %
and secondary 
BHs for a large distribution of binaries: 
at the initial separation $r_{\rm i}$, 
at the decoupling separation $r_{\rm decoup}$, %
and at the final separation of the post-Newtonian inspiral $r_{\rm GW}$, here taken as a proxy for the typical separations where sources becomes visible in LISA.
We assume fiducial parameters for the disk, BHs with masses and spins as in Fig.~\ref{F:Inspiral}, and initial spin misalignments $\theta_{1,2}$ that are isotropically distributed. The Bardeen-Petterson effect acts on the spins of BHs %
that are not initialized with a broken disk producing two peaks in the distribution of misalignments at $r = r_{\rm decoup}$.
The large, very localized peak at $\cos\theta_{1,2} \lesssim 1$ 
is composed of BHs initialized with %
angles smaller than the smallest possible $\theta_{\rm crit}$ evaluated at $r_{\rm i}$: the disk never breaks and alignment is very efficient. The broader peak at $\cos\theta_{1,2} \approx -0.20$ is composed of BHs whose spins experience partial alignment before the disk breaks, as they were initialized at angles larger than the largest possible $\theta_{\rm crit}$. The spins of BHs in the range $-0.20 \lesssim \cos\theta_{1,2} \lesssim 0.20$ are initialized in a region of the parameter space past criticality and %
do not evolve at all because of our assumptions. They therefore retain their initial isotropic spin orientations. This makes the smaller peak asymmetric across $\cos\theta_{1,2} \approx -0.20$ up to $r=r_{\rm decoup}$. After binaries decouple from the disk, GW emission drives the inspiral and erases the apparent asymmetry to produce the smaller peak in the distribution at $r=r_{\rm GW}$, which is instead roughly symmetric about $\cos\theta_2 = -0.20$. This implies that LISA might struggle to distinguish between binaries that align and then encounter criticality from binaries that are critical from the start.

\begin{figure}
\centering
\includegraphics[width=0.48\textwidth]{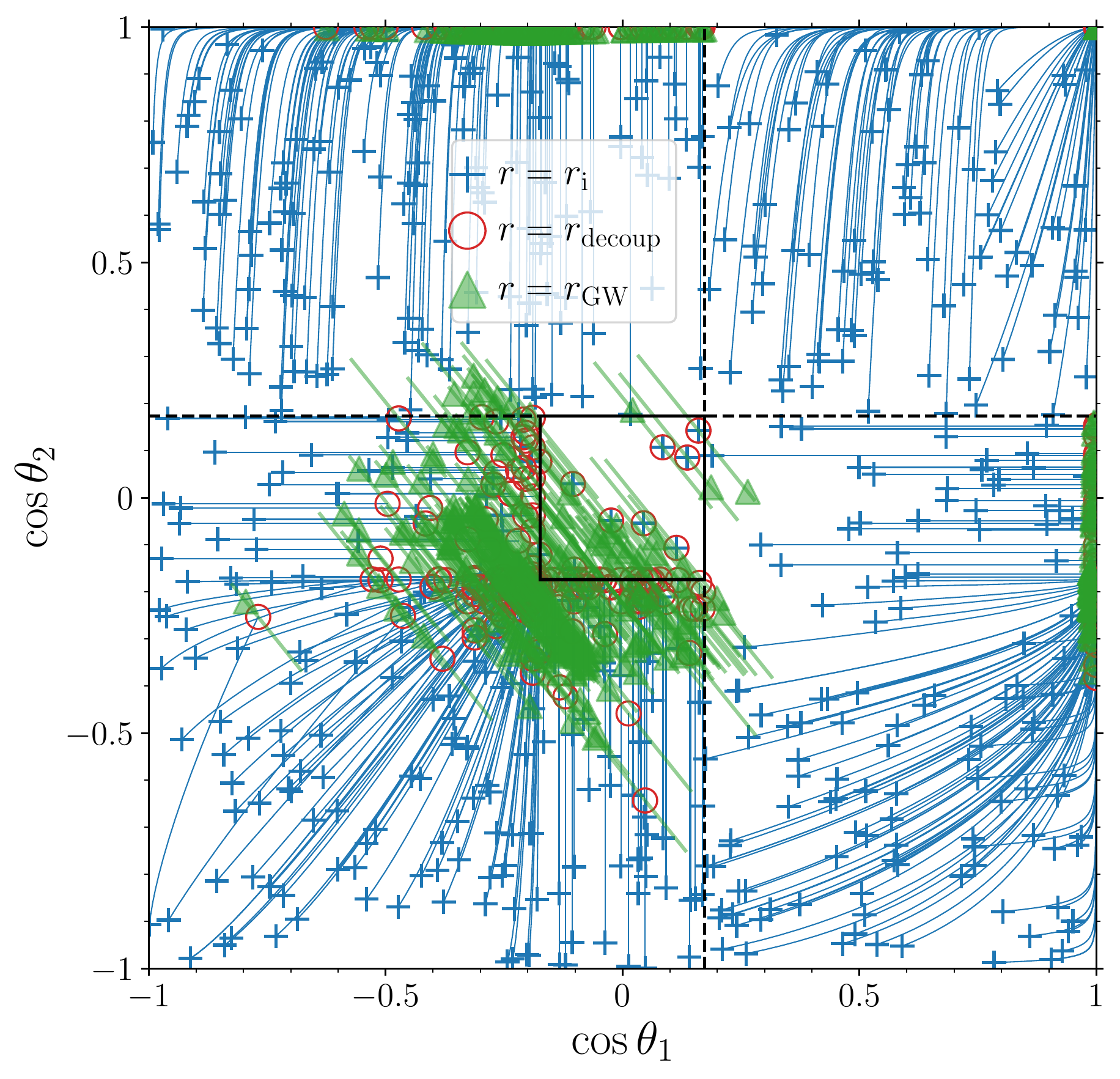}
\caption{The spin orientations of the primary $\cos\theta_1$ and secondary $\cos\theta_2$ BHs in binaries with total mass $M = 2\times10^7$ M$_{\odot}$, mass ratio $q = 0.8$, spin magnitudes $\chi_1 = \chi_2 = 0.1$, and fiducial values of the remaining parameters. Each binary is initialized at $r_{\rm i}$ [Eq.~(\ref{E:InitialSepScaling})] with isotropic spin directions shown by blue crosses, evolves through the phase of disk migration along a blue line until the decoupling separation $r_{\rm decoup}$ [Eq.~(\ref{E:DecoupSep})] shown by the red circles, and proceeds along contours of constant aligned effective spin $\chi_{\rm eff}$ shown by the green lines until $r_{\rm GW}$  [Eq.~(\ref{rgw})], denoted by the green triangles. The black box in the center contains binaries with spin orientations that do not evolve at all during the gas-driven phase as both BHs are initialized with broken disks. The black dashed lines correspond to the boundaries of the four subsets of binaries that either have both, only the primary or secondary, or neither spin aligned. 
} \label{F:ScatterExample}
\end{figure}

Figure~\ref{F:ScatterExample} shows the evolution of the same systems %
through the plane defined by the primary $\cos\theta_1$ and secondary $\cos\theta_2$ misalignments. As before, these BHs experience efficient alignment since $\omega_1 + \omega_2 > 1$, the secondary aligns more quickly than the primary since $\omega_2/\omega_1 > 1$. The fraction of binaries that begin the gas-driven migration with broken disks is enclosed by the black box in the center. The size of the box is set by the companion parameters $\kappa_{\rm i, 1} \simeq \kappa_{\rm i, 2} \sim 10^{-2}$. This results in four distinct subpopulations of binaries, each  originating from binaries that start from disjoint regions in the $(\cos\theta_1-\cos\theta_2)$ plane:
\begin{enumerate}[leftmargin=*]
 \item binaries with both spins aligned (top right); 
 \item binaries with only the primary-BH spin aligned (bottom right);
 \item binaries with only the secondary-BH spin aligned (top left);
 \item binaries with neither spin aligned (bottom left).
 \end{enumerate}
The boundaries of these regions can be easily computed from the condition of criticality and are shown with vertical and horizontal dashed black lines.

One can readily estimate the fraction of each subpopulation by leveraging the edges of the black box. For example, the fraction of binaries with two misaligned spins at $r = r_{\rm decoup}$ is the area defined by the upper and right edges of the black box extended along the dashed black lines toward the axes and divided by 4 (which is the total area of the plane).
Recall that the size of the black box depends on $\kappa_{1,\rm i}$, $\kappa_{2,\rm i}$, and $\alpha$. The fraction of binaries in this distribution that avoid alignment of both BHs (due to both BHs encountering critical angles) is $\simeq 30\%$, and the remaining three fractions are $\simeq 25\%$, $\simeq 25\%$, and $\simeq 20\%$ (the latter corresponds to the fraction with both spins aligned). 
As we 
explore %
in Sec.~\ref{subsec:ResGeneric}, %
the occurrence of four distinct subpopulations is not generic since both BHs may not experience efficient alignment, and the existence of each subpopulation is not guaranteed since it depends on the efficiency of alignment and the prevalence of the critical obliquity.

\begin{figure}
\centering
\includegraphics[width=0.48\textwidth]{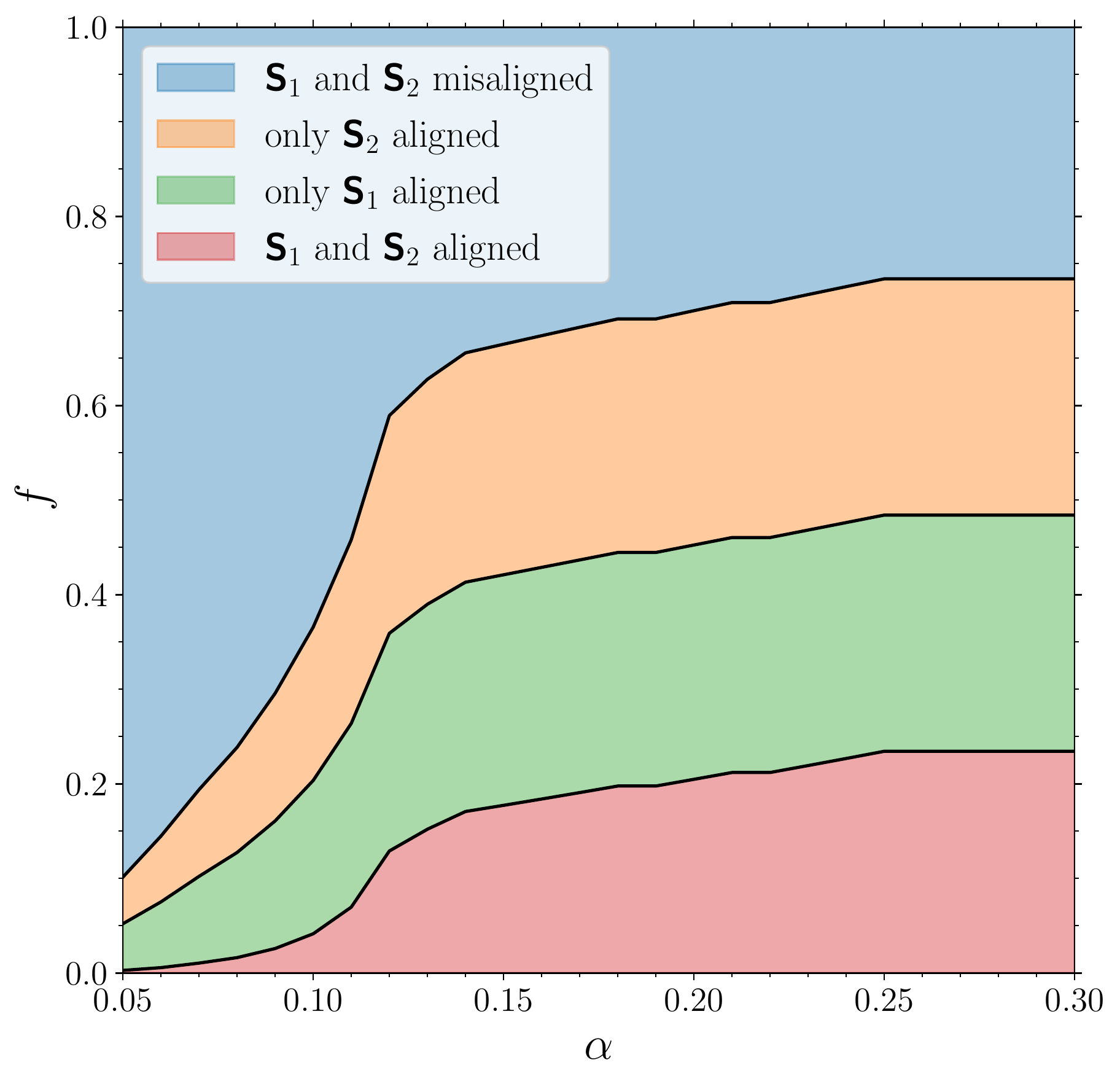}
\caption{The dependence of the relative contribution $f$ of each of the four binary subpopulations, defined by their spin orientations at the end of the disk migration inspiral, on the dimensionless viscosity parameter $\alpha$. All parameters but $\alpha$ are set as in Fig.~\ref{F:ScatterExample}. The red, green, orange, and blue regions correspond to the fractions of binaries with both spins aligned, with only the spin of the primary aligned, with only the spin of the secondary aligned, and with both spins misaligned, respectively.
} \label{F:FracVsAlpha}%
\end{figure}

In Fig.~\ref{F:FracVsAlpha} we show the fraction of binaries in each of these four regions  at the end of the gas-driven phase (i.e $r=r_{\rm decoup}$) as a function of the kinematic viscosity $\alpha$ while holding constant all other parameters that were chosen for Fig.~\ref{F:ScatterExample}. %
The fraction of binaries with both spins misaligned after disk migration 
is largest in the limit of small $\alpha$ since viscous disks are less likely to break. %
This is expected as the low-viscosity limit signals a break down 
in the $\alpha$-disk theory. %
As $\alpha$ increases, this fraction decreases monotonically while the remaining fractions of binaries with at least one aligned spin increase monotonically. The sharp feature at $\alpha \approx 0.1$ is consistent with the results of \cite{2020MNRAS.496.3060G}, see e.g. their Fig. 9. 
In the limit of large $\alpha$, all four fractions converge to $\approx 25\%$ because %
BHs are not initialized with broken disks: the area of the black box in Fig.~\ref{F:ScatterExample} tends to zero, thus dividing the $(\cos\theta_1-\cos\theta_2)$ plane into quadrants of equal area. 
Comparing with %
the contours of Fig.~\ref{F:Contours}, these fractions are largely insensitive to changes in the total mass and mass ratio, except in the limit of high total mass where $\kappa_{\rm i,1,2} \gg 1$ or in the limit of low total mass where $\omega_1 + \omega_2 \ll 1$. 
Both of these limits cause the fraction of binaries with two misaligned spins to approach unity, though for different reasons: the large-$\kappa$ limit causes the disks to break for all initial binary orientations while the low-$\omega$ limit causes the alignment to be very inefficient.

The subsequent evolution through the GW dominated phase of the inspiral is also shown in Fig.~\ref{F:ScatterExample}. When gravitational radiation begins, the spin orientations of BH binaries evolve along contours of constant aligned effective spin $\chi_{\rm eff}$ %
shown by the solid green lines. The extent of those contours, i.e., the largest and smallest values of $\theta_{1,2}$, depends on the separation and they are largest for the smallest separation, here taken to be $r_{\rm GW} = 10~GM/c^2$ %
shown by the green triangles. Only binaries that retained two significantly misaligned spins, due to encountering the critical obliquity during disk migration, experience significant variation on the radiation-reaction timescale. 
While  GW signals from these sources will exhibit significant spin-precession modulations \citep{2022PhRvD.106h4040D}, 
binaries with only one misaligned spin will exhibit a somewhat suppressed signature \citep{2020CQGra..37k5006O}. 
Binaries with both spins aligned, i.e. those in the upper-right corner at $r_{\rm decoup}$, will experience negligible spin precession, which in turn translates to a GW signal with a simpler morphology \citep{1994PhRvD..49.6274A}. These subpopulations of differing spin orientations and precession are a distinct signature of the Bardeen-Petterson effect in gas-rich hosts compared to a single population of generically precessing binaries in gas-poor hosts.

\subsection{Generic behavior during disk migration}
\label{subsec:ResGeneric}

\begin{figure*}
\centering
\includegraphics[width=\textwidth]{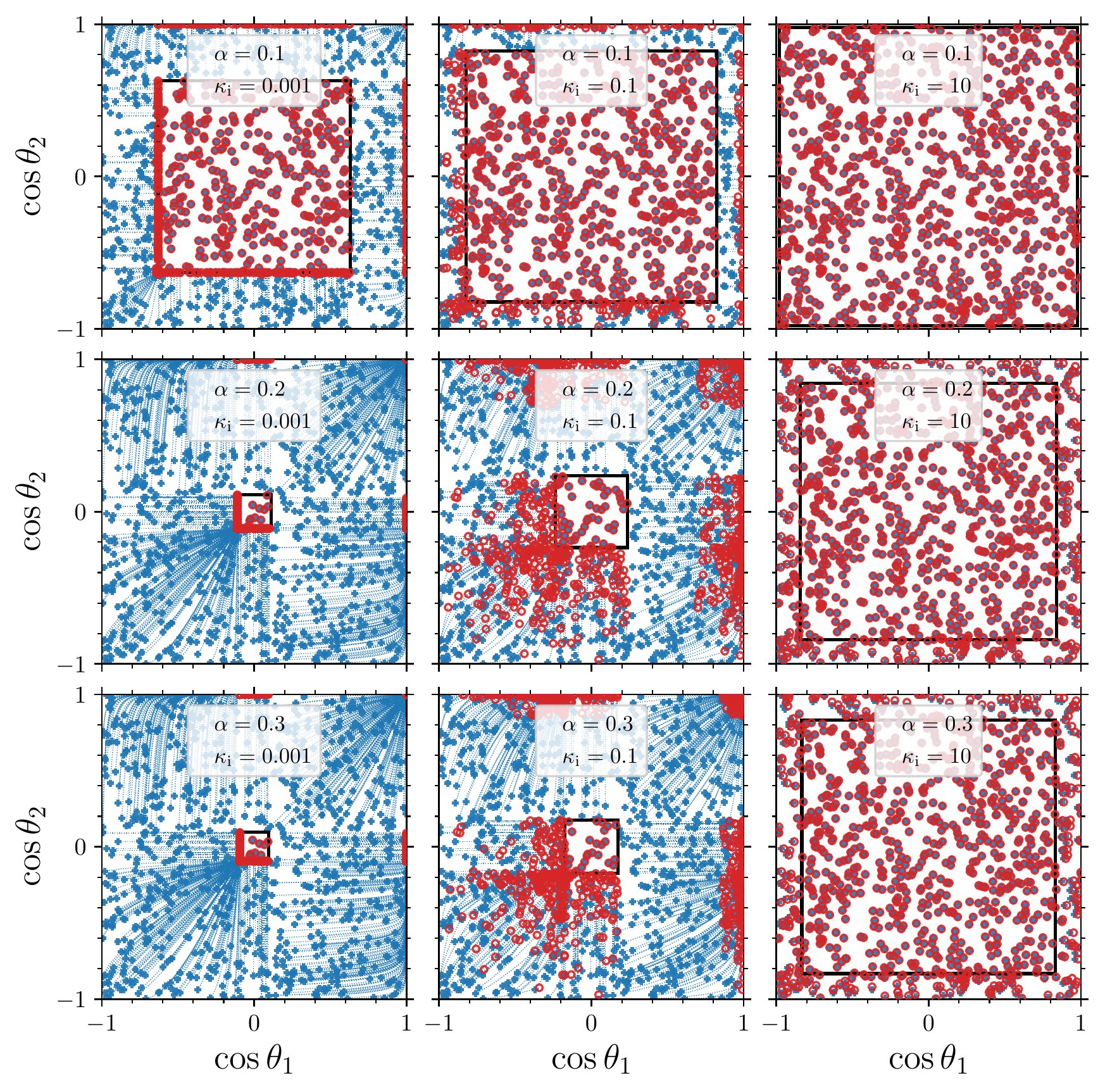}
\caption{
The evolution of the primary $\cos\theta_1$ and secondary $\cos\theta_2$ BH spin orientations through the phase of disk migration where binaries are initialized at $r_{\rm i}$ [Eq.~(\ref{E:InitialSepScaling})], shown by blue crosses, and evolve through the inspiral along the blue lines until the decoupling separation $r_{\rm decoup}$ [Eq.~(\ref{E:DecoupSep})], shown by the red circles. We assume here that the value of $\kappa$ at $r_{\rm i}$ is the same for the primary and secondary, i.e., $\kappa_{\rm i} \equiv \kappa_{\rm i, 1} = \kappa_{\rm i, 2}$, and that $\omega_1 = \omega_2 = 1$. Each panel shows the spin evolution for different values of $\kappa_{\rm i}$ and the viscosity $\alpha$: the left, middle, and right columns assume $\kappa_{\rm i} =$ 0.001, 0.1, and 10, respectively, and the top, middle, and bottom rows assume $\alpha =$ 0.1, 0.2, and 0.3, respectively. The black box centered in each panel contains binaries that are initialized %
with broken disks and
whose spin orientations do not evolve during this gas-driven phase. %
} \label{F:ScatterAlphaKappa}
\end{figure*}

\begin{figure*}
\centering
\includegraphics[width=\textwidth]{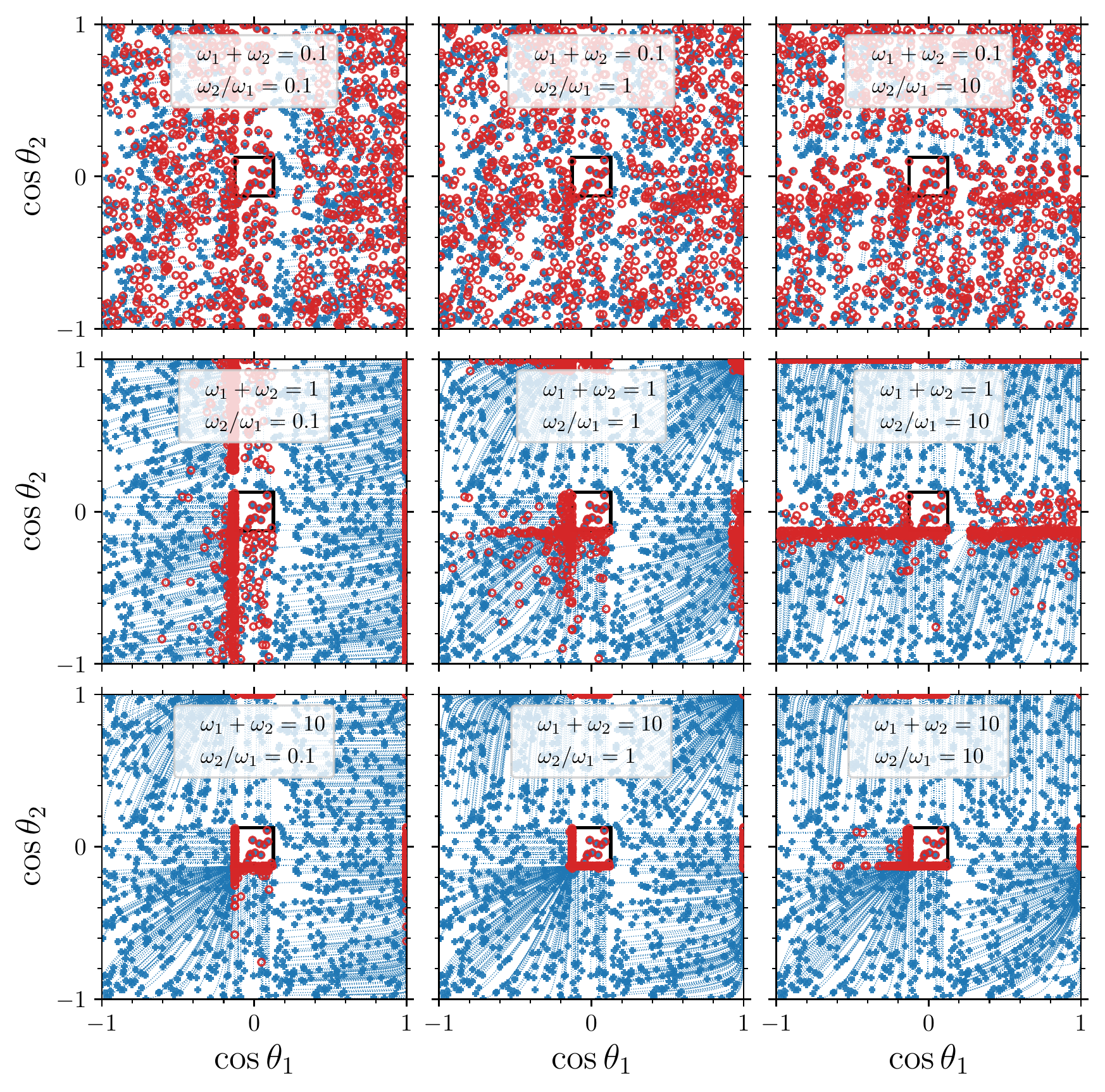}
\caption{
The evolution of the primary $\cos\theta_1$ and secondary $\cos\theta_2$ BH spin orientations through the phase of disk migration where binaries are initialized at $r_{\rm i}$ [Eq.~(\ref{E:InitialSepScaling})], shown by blue crosses, and evolve through the inspiral along the blue lines until the decoupling separation $r_{\rm decoup}$ [Eq.~(\ref{E:DecoupSep})], shown by the red circles. We assume here that the value of $\kappa$ at $r_{\rm i}$ is the same for the primary and secondary, $\kappa_{\rm i} =$ 0.01, and the viscosity is $\alpha =$ 0.2. Each panel shows the spin evolution for different values of the ratio and sum of $\omega_1$ and $\omega_2$: the left, middle, and right columns assume $\omega_1/\omega_2 =$ 0.1 1, 10, respectively, and the top, middle, and bottom rows assume $\omega_1 + \omega_2 =$ 0.1, 1, and 10, respectively. The black box centered in each panel contains binaries that are initialized %
with broken disks and
whose spin orientations do not evolve during this gas-driven phase. %
} \label{F:ScatterOmegas}%
\end{figure*}

We now attempt a broader exploration of the parameter space.
Despite the large number of parameters in our model, the phenomenology is greatly simplified because the evolutionary equation of the BH spin orientation under  disk-driven migration depends on only $\alpha$, $\beta$, $\gamma$, $\kappa_{{\rm i},1}$, $\kappa_{{\rm i},2}$, $\omega_1$, and $\omega_2$.  Other quantities enter Eqs.~(\ref{E:InitialSepScaling}) and (\ref{E:DecoupSep}) explicitly, but the precise prescription %
is irrelevant as long as $r_{\rm i} \gg r_{\rm decoup}$.
Furthermore, the spectral index of the viscosity profile $\beta$ was shown to have a %
minor effect on the disk-breaking process \citep{2020MNRAS.496.3060G} and, from our explorations, the index $\gamma$ mostly affects the path toward alignment and not the end state of the spins. %
We also noticed that setting  $\kappa_{{\rm i},1}= \kappa_{{\rm i},2} \equiv \kappa_{\rm i} $ still allows us to capture the broad phenomenology.  
This leaves four crucial parameters: $\alpha$, $\kappa_{{\rm i}}$, $\omega_{1}+\omega_2$, and $\omega_2/\omega_1$.

First, we will examine the effect of the parameters $\alpha$ and $\kappa_{\rm i}$. %
Together, these determine the possibility for a BH to encounter a critical configuration where the disk breaks. Then, we will examine the effect of the ``speed'' parameters $\omega_1$ and $\omega_2$ which govern the relative importance of spin alignment and inspiral for each BH. 

Figure~\ref{F:ScatterAlphaKappa} shows the evolution of binaries in the $(\cos\theta_1-\cos\theta_2)$ plane where we vary the values of $\alpha$ and $\kappa_{\rm i}$, assuming for simplicity that $\omega_1 = \omega_2 = 1$. In each panel, as in Fig.~\ref{F:ScatterExample}, binaries begin at $r_{\rm i}$, shown by the blue crosses, and evolve along the blue lines through the gas-driven inspiral which terminates at $r_{\rm decoup}$, shown by the red circles. The black box centered on %
$\theta_1=\theta_2=\pi/2$
 contains the binaries that are initialized with a disk that is already critical and thus whose spin orientations remain constant through the entire gas-driven inspiral. %
As $\kappa_{\rm i}$ increases or as $\alpha$ decreases, the fraction of those initially critical binaries (i.e. the area of the black box) increases preventing a larger proportion of binaries from aligning by the Bardeen-Petterson effect. 
For $\kappa_{\rm i} \geq 10$ %
as  in the panels in the right column, 
the black box contains nearly the entire plane for any value of $\alpha$ implying that the vast majority of disks are broken already at $r_{\rm i}$. Binaries initialized outside of the black box still experience very little alignment as this is only efficient at smaller $\kappa_{\rm i}$ for $\omega = 1$ \citep{2020MNRAS.496.3060G}. Comparing with the contours in Fig.~\ref{F:Contours}, binaries with $\kappa_{\rm i} \gtrsim 10$ have total mass $M \gtrsim 10^8$ M$_{\odot}$ which %
LISA is only sensitive to if they closer than $\ssim 40$ Gpc \citep{2017arXiv170200786A}. 
LISA will be more sensitive to binaries with total mass  %
$M \lesssim 10^7 M_{\odot}$, 
where $\kappa_{\rm i} \lesssim 0.01$ 
as in the panels in the left %
column of Fig.~\ref{F:ScatterAlphaKappa}. A significant fraction of these binaries experience spin alignment as long as the viscosity is not too low ($\alpha \gtrsim 0.1$), otherwise the broken disks occupy most of the parameter space for any value of $\kappa_{\rm i}$. Therefore, we generically find that BH binaries that evolve in gas-rich galactic hosts with low 
$\alpha$ or with large $\kappa_{\rm i}$ result in spin orientations that are %
largely
indistinguishable from the spins of binaries that evolve in gas-poor hosts. 

Sufficiently viscous accretion disks, i.e., $\alpha \gtrsim 0.1$ as in the two bottom rows of Fig.~\ref{F:ScatterAlphaKappa}, result in subpopulations of binaries defined by whether only one or both spins experience significant alignment, or both spins encounter a critical angle $\theta_{\rm crit}$ which ceases alignment. Due to the location of the critical obliquity in the parameter space, these subpopulations correspond to binaries that were initialized in precise regions of the spin-tilt plane (cf. Fig.~\ref{F:ScatterExample}). Depending on $\alpha$ and $\kappa_{\rm i}$, disk migration compresses the spin orientations of these binaries into regions that are either very compact (cf. the panels in the first column, second and third rows) or more locally dispersed (cf. panels in the second column, second and third rows). The difference here is the value of $\kappa_{\rm i}$, where binaries in the first column have more time to align before criticality than those in the second column (though if we had assumed larger values of $\omega_{1,2}$ both columns would result in highly aligned systems and compact regions). The compact and localized regions of binaries in the $(\cos\theta_1-\cos\theta_2)$ plane are a distinct signature of the Bardeen-Petterson effect acting in supermassive BH binaries. 
Equivalently, one can view the complementary regions that are vacant of binaries as indicating the presence of spin alignment, with stricter vacancy corresponding to more efficient alignment.

Next, in Fig.~\ref{F:ScatterOmegas} we vary the ratio and sum of $\omega_1$ and $\omega_2$ while fixing $\alpha = 0.2$ and $\kappa_{\rm i} = 0.01$. The ratio $\omega_2/\omega_1$ governs the relative speed of alignment with respect to each BH and the sum $\omega_1 + \omega_2$ governs the relative speed of alignment with respect to the binary inspiral.
For $\omega_2/\omega_1 \ll 1$ ($\omega_2/\omega_1 \gg 1$), as in the panels in the left (right) column, the primary (secondary) BH aligns more quickly than its companion. When $\omega_1 + \omega_2 \ll 1$, as in the panels in the top row, the alignment of each BH is slow relative to the binary inspiral, implying that the spin orientations will not evolve significantly during the gas-driven phase. Instead, when $\omega_1 + \omega_2 \gtrsim 1$, as in the panels in the middle and bottom rows, at least one of the BHs aligns quickly relative to the inspiral, implying that the binaries move significantly through the $(\cos\theta_1-\cos\theta_2)$ plane. The spin orientations of binaries in the panels of the middle row accumulate in horizontally or vertically oriented regions depending on $\omega_2/\omega_1$, i.e., on which BH aligns more quickly. In the bottom row of panels, the value $\omega_1 + \omega_2 \gg 1$ causes both BHs to experience efficient alignment regardless of the value of $\omega_2/\omega_1$. The combinations of $\omega_2/\omega_1$ and $\omega_1 + \omega_2 \gtrsim 1$ shown in the panels of the bottom row result in analogous subpopulations created by the Bardeen-Petterson effect as in the left column and two bottom rows of Fig.~\ref{F:ScatterAlphaKappa}.

In Figs. \ref{F:ScatterAlphaKappa} and \ref{F:ScatterOmegas} we assumed $\kappa_{\rm i, 1} = \kappa_{\rm i, 2}$ for simplicity, which yields a square critical region in the ($\cos\theta_1 - \cos\theta_2$) place as $\kappa_{\rm i, 1}$ determines the width and $\kappa_{\rm i, 2}$ determines the height of the black box.  %
This provides a statistically equal number of primary and secondary BHs initialized with broken disks.
In full generality, these quantities are not necessarily equal implying that the black box need not be a square. A non-square region is possible, for examples, for low mass ratio systems and unequal spin systems [cf. Fig.~\ref{F:Contours} and Eq.~(\ref{E:kappa})], which would result in either the primary or the secondary BH to be more likely initialized at a critical disk configuration.

\section{Conclusions and Discussion}
\label{sec:Discussion}

The mergers of supermassive binary BHs will be a key target for the future LISA detector, offering a unique opportunity to probe unknown astrophysical processes behind their formation and evolution. In this work, we systematically explored the consequences of the Bardeen-Petterson effect on the spin orientations of BH binaries in gas-rich hosts and demonstrated how its imprint on the spins could aid in distinguishing these binaries from those that evolve in gas-poor hosts. Improving upon previous work on the topic \citep{2007ApJ...661L.147B,2013MNRAS.429L..30L,2013ApJ...774...43M,2015MNRAS.451.3941G} we consider the impact of the critical obliquity ---a specific region in the parameter space where the disk breaks and spin alignment is halted  \citep{2014MNRAS.441.1408T,2020MNRAS.496.3060G,2022MNRAS.509.5608N}.

We find that disk breaking and the subsequent suppression of spin alignment introduces degeneracies between the spins of binaries that evolve in gas-poor hosts and those that evolve in gas-rich hosts. In particular, this is most relevant for binaries with either:
\begin{enumerate}[leftmargin=*]
    \item low viscosity $\alpha \lesssim 0.1$ due to an enhanced likelihood of disk breaking, %
    \item very high total mass $\gtrsim 10^8$ M$_{\odot}$ (i.e., large $\kappa_{\rm i} \gtrsim 1$) due to beginning disk migration past a critical obliquity, or
    \item very low total mass $\lesssim 10^6$ M$_{\odot}$ (i.e., small $\omega_1 + \omega_2 \lesssim 0.1$) due to inefficient alignment. 
\end{enumerate}
LISA will likely measure supermassive BHs with masses $\sim 10^4 - 10^8$ M$_{\odot}$ creating a ``Goldilocks zone'' for binaries with $M\sim 10^6 - 10^8$ M$_{\odot}$ to have aligned spins $\emph{and}$ to be observable by LISA. 
{From our extended investigation, including several runs not reported here for clarity, we find that this optimal region for alignment and observability is generic across the parameter space of our model. %
The mass ratio $q$ and the ratio of the dimensionless spin magnitudes determine the relative speed of alignment between the BHs in a binary [cf. Eq.~(\ref{E:OmegaRatio})], implying that highly asymmetric masses or spins cause a preference for larger misalignment of one BH.} 

When Bardeen-Petterson alignment is efficient, i.e  $\omega_1+\omega_2\gtrsim 1$ or total mass $M \gtrsim 10^6$ M$_{\odot}$ for fiducial disk parameters, a distribution of binaries with initially isotropic spin orientations evolve into distinct subpopulations defined by whether neither, both, or only one of the two spins are aligned by the time that GW emission begins to dominate the inspiral. 
A signature of efficient alignment is the occupancy of these subpopulations in highly localized, compact regions of the ($\cos\theta_1-\cos\theta_2$) plane.
The number of possible subpopulations depends on the relative speed of alignment between the BHs $\omega_2/\omega_1$. For examples, binaries with equal spin magnitudes and mass ratio $q \gtrsim 0.5$ yield four subpopulations as both BHs experience efficient alignment %
whereas binaries with very asymmetric $q$ yield only two subpopulations as only one BH experiences efficient alignment, i.e., see the panels in the left and right columns and middle row of Fig.~\ref{F:ScatterOmegas}. The relative contribution of each subpopulation then crucially depends on the kinematic viscosity $\alpha$ (Fig.~\ref{F:FracVsAlpha}). %

We conclude that measurements of aligned spins by LISA can be considered a smoking-gun signature of the Bardeen-Petterson effect, consistent with previous work \citep{2013ApJ...774...43M,2015MNRAS.451.3941G}. This implies a strong correlation between the prevalence of alignment processes during the binary inspiral in gas-rich hosts and the directly measurable spin precession exhibited by the binary through the LISA detection band prior to merger. Although beyond the scope of this work, our results suggest that a binary will experience suppressed spin precession if the Bardeen-Petterson effect efficiently aligns the spin of at least one BH with the binary orbital angular momentum. Meanwhile, a sufficiently large spin misalignment of the companion BH indicates the presence of the critical obliquity that caused the accretion disk to break. A binary with two highly misaligned BH spins indicates that both accretion disk broke due to the critical obliquity. This suggests that disk breaking should be correlated with significant spin precession. Thus, LISA has the capability to probe not only the process of alignment, but also the possibility of disk breaking in gas-rich astrophysical environments. At the same time, however, largely misaligned spins are also predicted for gas-poor systems, suggesting a partial degeneracy with the models explored here. 
Further investigations in the context of statistical model selections are ongoing.

Beside GW measurements with LISA, recoils are another interesting observable~\citep{2012AdAst2012E..14K} that could potentially constrain our models.
Merging BH binaries receive recoil velocities as large as $\ssim 5000$ km/s as a result of linear momentum conservation from anisotropic GW emission, with the largest kicks predicted for sources with highly misaligned spins  \citep{2007PhRvL..98w1101G,2007ApJ...659L...5C}. Tracking the spin evolution of BHs %
is thus crucial for predicting the post-merger proper velocity and hence the occurence of off-nuclear quasars.   While BH recoils can even exceed the escape velocity of the most massive galaxies in the Universe \citep{2004ApJ...607L...9M,2015MNRAS.446...38G},  %
 it was previously claimed that systems in gas-rich galaxies are unlikely to be ejected precisely because of disk accretion (e.g. \citealt{2012PhRvD..85h4015L,2012MNRAS.423.2533B}). 
Our results imply that suppressed alignment from disk breaking complicates this expectation. %

We argue our model encapsulates the essential ingredients of the alignment of binary BH spin orientations during disk migration. At the same time, there are several caveats that require further investigation. Although uncertain, prior to disk migration the BH spin orientations may experience alignment due to %
gaseous dynamical friction \citep{2010MNRAS.402..682D}. %
Our predictions rely on an effective fluid disk theory where the viscosity is encapsulated into the \cite{1973A&A....24..337S} $\alpha$ parameter. Developing calibration/fitting strategies using magnetohydrodynamics simulations (for instance, note the recent works by \citealt{2021MNRAS.507..983L,2022arXiv220103085M,2022arXiv221010053K} which tackle disk breaking) or providing sub-grid prescriptions for large-scale simulations (see e.g. \citealt{2018MNRAS.477.3807F}) offers an interesting avenue for future work.
The dependence of our results on $H/R$ is also expected to be important, cf. the steep dependence $\kappa \propto (H/R)^{-6}$ in Eq.~(\ref{E:kappa}). This deserves a careful investigation, including the re-examination of some of our assumptions such as the adopted differential-accretion prescription and the relative temperature of the circumbinary and secondary disks.
Perhaps most importantly, it is unclear how the evolution proceeds after the critical obliquity is reached and the disk breaks. 
In principle, 
some angular momentum can still be transferred through the system such that spin alignment is suppressed and not stopped completely as assumed here. This implies that, when alignment is efficient in our Figs.~\ref{F:ScatterExample}, \ref{F:ScatterAlphaKappa}, and \ref{F:ScatterOmegas}, BHs may not be so tightly localized and instead smear out along the borders and in the center of the ($\cos\theta_1 - \cos\theta_2$) plane. %
Our model is valid in a quasi-adiabatic approximation where the viscous, alignment, and accretion timescales are well separated, such that we can neglect changes to the mass and spin magnitudes of the accreting BHs. Relaxing this assumption might introduce interesting dependencies between those quantities and the spin directions. Lastly, a more consistent treatment of the binary evolution would evolve the spin orientations simultaneously under disk migration and GW emission, at least in the transition region where $r \sim r_{\rm decoup}$.

LISA is expected to measure several BH-binary spin orientations with a conservative accuracy of $\Delta \theta_{i} \lesssim 10^\circ$ \citep{2016PhRvD..93b4003K}, opening for the concrete possibility of probing fine details of the warped-disk dynamics such as the Bardeen-Petterson effect and the critical obliquity. A detailed investigation of the LISA signals predicted by our models as well as the instrumental capability to measure the underlying model parameters is left to future work.

\section*{Acknowledgements}
We thank Rebecca Nealon, Massimo Dotti, Alberto Sesana, Roberto Cotesta, Giovanni Rosotti, and Enrico Ragusa for discussions. N.S. and D.G. are supported by Leverhulme Trust Grant No. RPG-2019-350, European Union's H2020 ERC Starting Grant No. 945155--GWmining, and Cariplo Foundation Grant No. 2021-0555. Computational work was performed at CINECA with allocations through INFN, Bicocca, and ISCRA project HP10BEQ9JB.

\section*{Data Availability}
The data underlying this article will be shared on reasonable request to the correspondence author.

\bibliographystyle{mnras_tex_edited}
\bibliography{BPlisa}

\end{document}